\documentclass[10pt,pre,aps,twocolumn,superscriptaddress,floatfix,showpacs]{revtex4}
\usepackage{graphicx}
\usepackage{amssymb}
\usepackage{bm}
\newcommand{\dn}{\downarrow}
\newcommand{\up}{\uparrow}

\begin{document}

\title{Plaquette Renormalization Scheme for Tensor Network States}

\author{Ling Wang} 
\affiliation{Department of Physics, Boston University, 590 Commonwealth Avenue, Boston, Massachusetts 02215}
\affiliation{University of Vienna, Faculty of Physics, Boltzmanngasse 5, 1090 Vienna, Austria}

\author{Ying-Jer Kao} 
\affiliation{Department of Physics, National Taiwan University, Taipei, Taiwan 106}

\author{Anders W. Sandvik} 
\affiliation{Department of Physics, Boston University, 590 Commonwealth Avenue, Boston, Massachusetts 02215}
\affiliation{Department of Physics, National Taiwan University, Taipei, Taiwan 106}

\date{\today}

\begin{abstract}
We present a method for contracting a square-lattice tensor network in two dimensions, based on auxiliary tensors accomplishing
successive truncations (renormalization) of 8-index tensors for $2\times 2$ plaquettes into 4-index tensors. Since all approximations are
done on the wave function (which also can be interpreted in terms of different kind of tensor network), the scheme is variational, and, thus, 
the tensors can be optimized by minimizing the energy. Test results for the quantum phase transition of the transverse-field Ising model 
confirm that even the smallest possible tensors (two values for each tensor index at each renormalization level) produce much better 
results than the simple product (mean-field) state. 
\end{abstract}

\pacs{02.70.Ss, 75.10.Jm, 75.40.Mg, 75.40.Cx}

\maketitle

\section{Introduction}
\label{sec1}

\begin{figure}[b]
\includegraphics[width=6cm, clip]{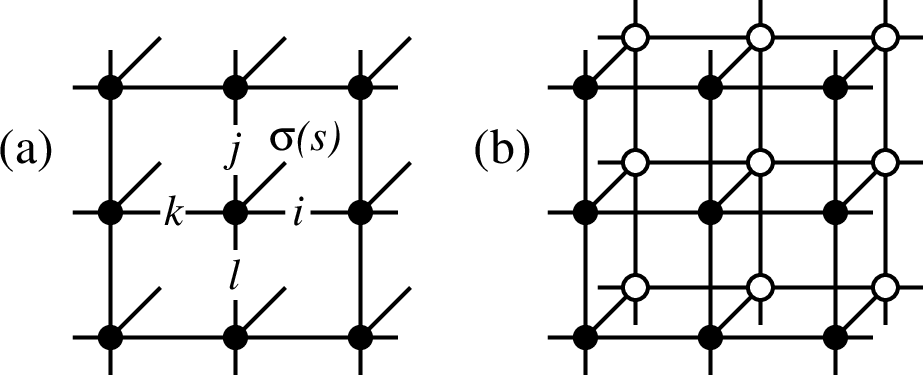}
\caption{Tensor networks on the 2D square lattice. (a) For each site $s$ there is a 4-index tensor $T^s_{ijkl}(\sigma_s)$, where $\sigma_s$ is the physical index 
(here a spin $\sigma_s =\pm 1$) and $i,j,k,l \in \{ 1,\ldots,m\}$. The wave function $\Psi(\sigma_1,\ldots,\sigma_N)$ is the $N$-tensor product (contraction) 
defined as a summation over all shared indices on the lattice bonds. (b) The norm $\langle \Psi|\Psi\rangle$ is obtained by contracting also over the physical 
indices.}
\label{net}
\end{figure}

Tensor network states (TNSs) \cite{verstrate1,nishino,verstrate2,vidal1} are emerging as a promising route toward unbiased modeling of challenging quantum many-body 
systems, such as frustrated spins. These correlated states are higher-dimensional generalizations of the matrix product states (MPSs) \cite{ostlund,verstrate3}
that are implicitly produced in density matrix renormalization group (DMRG) calculations \cite{white,schollwock} and are known to faithfully represent ground states of 
one-dimensional (1D) hamiltonians with short-range interactions \cite{vidal2}. The matrix size $m$ has to increase at most polynomially with the system size $N$, which 
underlies the success of the DMRG method in 1D. For 2D and 3D systems, correlations are not reproduced properly by MPSs \cite{verstrate2}, due to the inherently 
1D nature of the local quantum entanglement in these states (although improved schemes \cite{schuch,sandvik1} can restore 2D or 3D uniformity), and $m$ then has to grow 
exponentially with $N$. In the TNSs, the matrices are replaced by tensors of rank corresponding to the coordination number of the lattice, e.g., on a 2D square lattice 
the tensors $T^s_{ijkl}(\sigma_s)$ for each site $s$ have four indices, in addition to their physical index (here the $z$-component $\sigma_s$ of a spin), as illustrated 
in Fig.~\ref{net}. Contracting over the ``bond indices'' ($i,j,k,l$) gives the wave function coefficient for given spin state $\sigma_1,\ldots,\sigma_N$.

While it is believed, based on entanglement 
entropy considerations \cite{verstrate2}, that TNSs can represent ground states of short-range 2D and 3D hamiltonians, a serious problem in practice is that contracting 
the tensors is, in general, a problem which scales exponentially in the system size and the dimension of the tensors. To overcome this challenge, approximate ways to 
compute the contraction have been proposed \cite{verstrate1,gu,xiang}. Another approach is to use tree-tensor networks \cite{vidal1}, or more sophisticated extensions of 
these \cite{vidal2}, which can be efficiently contracted. Promising results based on TNSs have already been reported for several quantum spin models, but further reduction 
of the computational complexity, while maintaining the ability of the TNSs to properly account for entanglement, will still be necessary before the most challenging 
systems can be studied reliably.

In this paper, we introduce a plaquette renormalization scheme for 2D TNSs inspired by a method introduced by Levin and Nave \cite{levin} in the context
of classical Ising models. They suggested to replace the effective tensors for $2\times 2$ plaquettes on the square lattice, which have $m^8$ elements, by some 
``renormalized'' tensors with $m_{\rm cut}^4$ elements, as illustrated in Fig.~\ref{pren}(a). This can be done exactly if all tensor elements are kept, with 
$m_{\rm cut}=m^2$, as this corresponds just to a regrouping of the tensor indices. The idea is that the scheme may provide a good approximation even if the tensors 
are drastically truncated, e.g., with $m_{\rm cut} \propto m$. For a square $L\times L$ lattice, the new tensors describing $2\times 2$ spins should be contracted on a 
new lattice of length $L/2$. If the original $L$ is a power of $2$, this decimation can be continued until there is a single tensor left, which is then contracted 
with itself (under periodic boundary conditions) to give the wave function. The question is then how to accomplish the tensor truncation in a way which preserves 
the quantum state in the best way. 

We here propose a renormalization procedure based on auxiliary tensors introduced at the level of the wave function, as illustrated in
Fig.~\ref{trg}, in contrast to recent schemes \cite{gu,xiang} that apply singular-value decompositions on the ``double tensor'' product obtained when the 
physical indices have been traced out. Our approach is strictly variational, and can be used also in combination with Monte Carlo sampling of the spins 
\cite{sandvik2}. Here we carry out calculations on the 2D transverse-field Ising model, performing the contraction over the spins exactly.

The outline of the paper is as follows: In Sec.~\ref{sec2} we first discuss tensor renormalization schemes in general and then provide details of our variant. 
We test the method by studying the quantum phase transition in the 2D transverse-field Ising model in Sec.~\ref{sec3}. In Sec.~\ref{sec4} we summarize and discuss 
several aspects of the method and its possible future extensions and applications.

\section{Tensor renormalization}
\label{sec2}

\begin{figure}[t]
\includegraphics[width=6.5cm, clip]{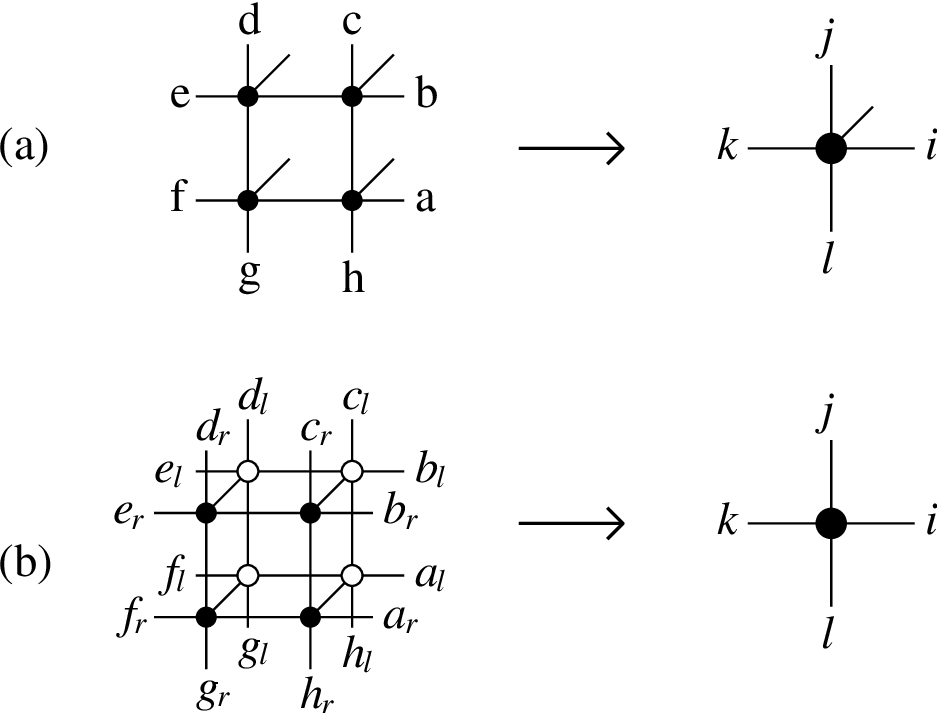}
\caption{(a) Truncation (renormalization) of an 8-index plaquette tensor into an effective 4-index tensor. The indices $a,\ldots,h \in \{1,\ldots,m\}$ 
and $i,j,k,l \in \{ 1,\ldots,m_{\rm cut}\}$. The diagonal lines indicate physical indices, which for an $S=1/2$ spin system can take two values before 
the renormalization and $16$ values after (and is further multiplied by 16 after each successive renormalization). In (b) the physical indices are traced out 
first, leading to double tensors. To stay with the same level of truncation as in (a), the indices after the renormalization should then take values 
$i,j,k,l \in \{ 1,\ldots,m_{\rm cut}^2\}$.}
\label{pren}
\end{figure}

In order to compute physical expectation values based on a TNS, one has to contract the tensors of a bra and ket state over their physical (e.g., spin) indices 
in addition to the bond indices of the tensors. The full contraction of such a 2D double-tensor network, illustrated in Figs.~\ref{net}(a) and \ref{net}(b), is the 
norm $\langle \Psi|\Psi\rangle$. A matrix element $\langle \Psi|A|\Psi\rangle$ of some operator involving one or several sites can be treated in a similar way. 
In practice, in most TNS approaches, one would first construct the double tensors
\begin{equation}
D^s_{abcd}= \sum_{\sigma_s=\up,\dn}T^{s*}_{i_2j_2k_2l_2}(\sigma_s)T^s_{i_1j_1k_1l_1}(\sigma_s),
\label{decont}
\end{equation}
where the labels $a,b,c,d$ is a suitable combination of the indices of the bra ($T^{s*}$) and ket ($T^s$) tensors, 
i.e., $a=i_1+m(i_2-1)$, etc. Similar tensors are constructed for
the sites at which operators act in a local expectation value. The full contraction of the double tensors $D$, which have bond dimensions $m^2$, is then carried out 
in some approximate way. A plaquette renormalization in the Levin-Nave scheme with double tensors on the square lattice is depicted in Fig.~\ref{pren}(b). This scheme is 
exact if the full tensors (with bond dimension $m^4$) are kept, but this is infeasible in practice. The scheme may be a good approximation for some judiciously 
chosen truncated (renormalized) tensor, but how to find the optimal way to construct it is an open question. Gu, Levin, and Wen implemented a singular value 
decomposition (SVD) scheme \cite{gu} for the double tensors, and a similar method was proposed by Jiang, Weng, and Xiang \cite{xiang}. 

\begin{figure}
\includegraphics[width=4.cm, clip]{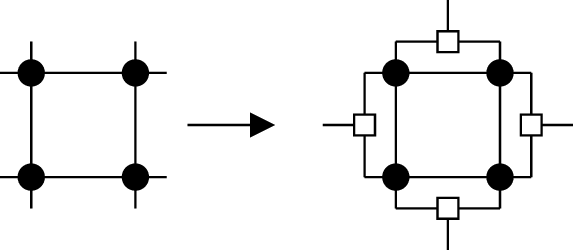}
\caption{Renormalization of an 8-index plaquette tensor using auxiliary 3-index tensors. The solid circles denote either the original tensors 
$T=T^0$, which depend on the physical spins (which are not shown here), or tensors $T^n$ arising after $n$ renormalization steps have been carried out 
(and depend on the $2^{2n}$ spins within the block). The squares denote 3-index tensors $S^n$ by which the external indices are decimated by contracting 
common $T^{n-1}$ and $S^n$ indices. The remaining four free indices take values $1,\ldots,m_{\rm cut}$.}
\label{trg}
\end{figure}

\begin{figure}
\includegraphics[width=5.5cm, clip]{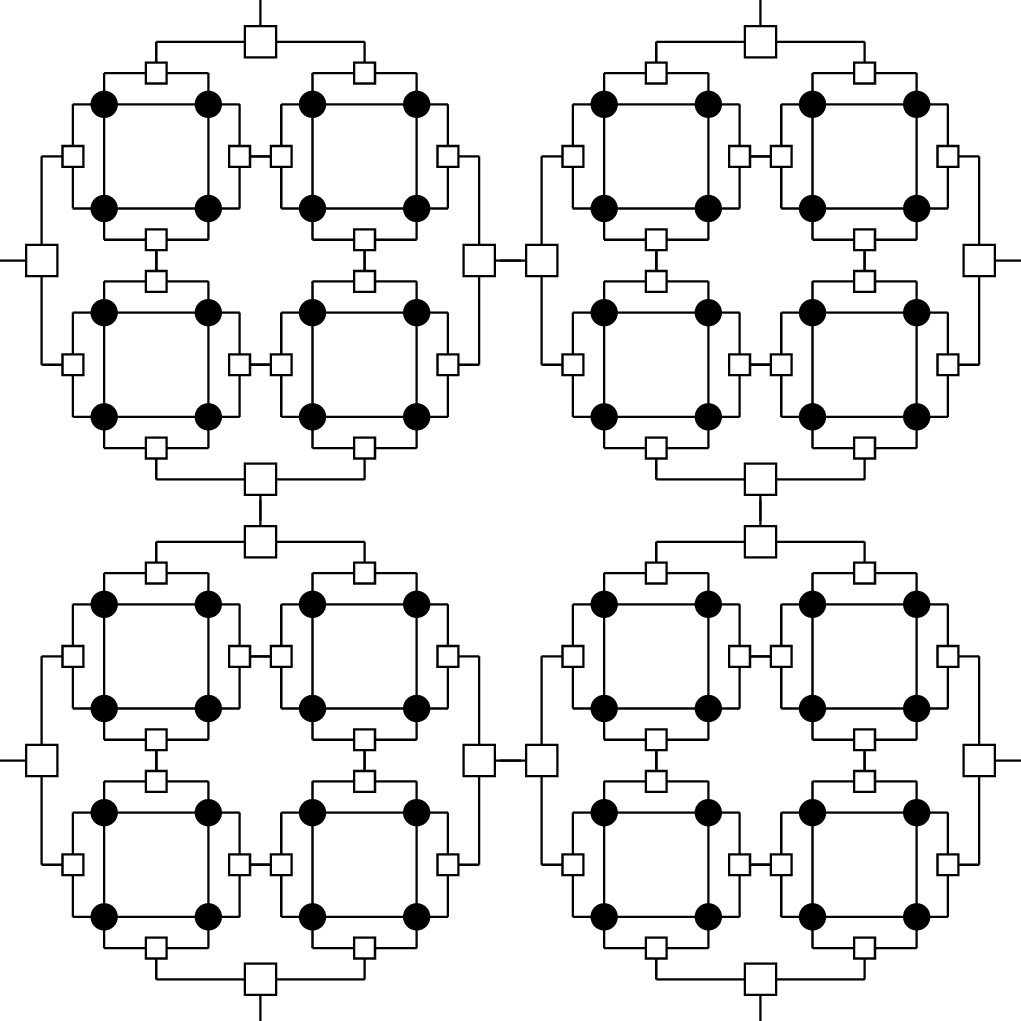}
\caption{The effective reduced tensor network for the wave function on an $8 \times 8$ lattice. Here there are two sets of $S$ tensors; $S^1$ and $S^2$, 
denoted by the smaller and larger squares. The solid circles correspond to the original, spin dependent tensor $T^0$. At the lowest level (the black circles),
there is also a spin index of the tensors which is not indicated here.}
\label{net2}
\end{figure}

In our scheme, the renormalization is instead accomplished with the aid of auxiliary 3-index tensors $S^n_{abc}$ in the wave function, which transform and truncate 
pairs of indices of the plaquette tensors, as shown in Fig.~\ref{trg}. A sequence of plaquette renormalizations, $n=1,2,\ldots$, effectively corresponds to a different
kind of tensor network, which is illustrated in the case of the wave function on an $8 \times 8$ lattice in Fig.~\ref{net2}. An advantage of working with the wave
function, instead of the contracted double-tensor network, is that the method remains variational, regardless of how the $S$-tensors are chosen. This
is not strictly the case when approximations are done when contracting the double tensors.

Here we will apply this scheme to the transverse-field Ising model (which has become the bench-mark of choice for initial tests within TNS approaches);
\begin{equation}
H = -J \sum_{\langle ij\rangle} \sigma^z_i\sigma^z_j -h \sum_{i} \sigma^x_i,
\label{ham}
\end{equation}
where $\sigma^x_i$ and $\sigma^z_i$ are standard Pauli matrices and $\langle ij\rangle$ denotes nearest-neighbor site pairs on a 2D 
square $L\times L$ lattice with periodic boundary conditions. The ground-state wave function of this system is translationally invariant and positive-definite. 
We can then take the original tensors to be site-independent, i.e., there are just two tensors $T_{ijkl}(\sigma_s=\pm 1)$ and the tensors $S^n$ are the same for 
all plaquettes on a given renormalization level $n=1,...,\log_{\rm 2}(L)-1$. At the last level, four tensors remain (see Fig.~\ref{net2}), which we contract 
directly. The problem is now to optimize the elements of the $T$ and $S$ tensors, to minimize the energy.  Before proceeding to calculations, several 
comments are in order. 

It is clear from Fig.~\ref{net2} that the plaquette renormalization (no matter how it is accomplished) breaks translational symmetry, which, in the optimized state, 
should gradually be restored with increasing $m$. A way to restore the symmetry for finite $m_{\rm cut}$ is to sum over all (here four) symmetrically non-equivalent ways of 
arranging the plaquettes on the lattice at each level \cite{sandvik1}. In a similar way, one can also ensure that the wave function is symmetric under other lattice 
transformations (rotations and reflections) for arbitrary $T$ and $S^n$ (as an alternative to enforcing these symmetry in the individual tensors), and spin-inversion 
symmetry can be implemented in a similar way. These symmetrization procedures (which also enable studies of states with different quantum-numbers of the symmetry operators) 
can be carried out if the spins are sampled using Monte Carlo simulations \cite{sandvik2}, but cannot be easily used with the double-tensor network. Here, in this initial 
test of the scheme, we will trace out the spins exactly and work with the double tensors. When minimizing the energy, it is then important to calculate the full, 
translationally averaged energy, not just the site and bond energies on a single plaquette (as is done in the SVD scheme of Gu {\it et al.} \cite{gu}), in order to maintain 
the variational property of the scheme. No approximations are made when contracting the effective renormalized tensor network, also when using the double tensors. In contrast,
the SVD applied to the double-tensor network \cite{gu} introduces an approximation due to which the calculated energy becomes non-variational. This can cause 
problems when optimizing the tensors. Note that in our double-tensor approach, there is a pair of equal $S$ tensors for each plaquette edge (one from $\langle \Psi|$ 
and one from $|\Psi\rangle$) and one cannot combine these into arbitrary tensors, which would make the scheme non-variational. 

\begin{figure}
\includegraphics[width=8.25cm, clip]{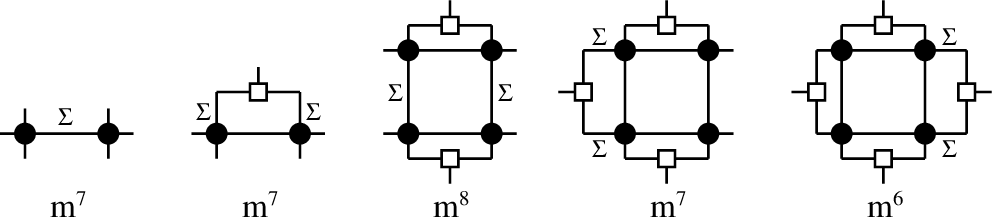}
\caption{Steps in the contraction of the renormalized plaquette tensor of Fig.~\ref{trg}. Partial contractions are constructed from left to right.
The summations carried out are indicated by $\Sigma$ and the scaling of each step with $m$ is shown beneath the diagrams. Each summation and free
index contributes a factor $m$.}
\label{cont4}
\end{figure}

The renormalization of a plaquette according to Fig.~\ref{trg} requires 12 internal index summations for each combination of the four external indices of the 
renormalized tensor, which can be carried out with $\propto m^8$ operations (assuming $m_{\rm cut}=m$, which we will use here). One example of a sequence of contractions 
giving this scaling is shown in Fig.~\ref{cont4}. When working with double tensors (as we do here, contracting first the spin dependent tensors at the lowest level
over the spin indices), the scaling becomes $m^{16}$, since all external and internal indices should then take $m_{\rm cut}^2=m^2$ values. Note that the structure
with $S$ tensors renormalizing the bra and ket state (the same for the bra and the ket) has to be preserved at each level, i.e., it is not correct to just compute 
the double tensor with the lowest-level spin dependent tensors and after that use a single $S$ tensor on each level to renormalize the double-tensor network. Such 
a scheme would not be variational.

The derivatives of the energy 
with respect to all the tensor elements, which we use to minimize the energy, can be evaluated in $\propto m^{12}$ operations ($m^{24}$ with the double tensors), using a 
chain-rule procedure carried out along with the renormalization steps. 
We need the derivative of $E$ with respect to the elements of the spin-dependent tensors $T^0(\sigma)$, as well as the renormalization tensors $S^n$ for all levels 
$n=1,\ldots,n_{\rm max}$. The procedure is very similar for both $T$ and $S$ and we write down expressions only for the former case. To simplify the notation we suppress the 
spin dependence of the tensors $T^0(\sigma)$ (i.e., there are equations for both $\sigma=\pm 1$). We first note that we can write the derivative needed in terms of the 
renormalized tensor at any level $n$ as
\begin{equation}
  \nonumber
  \frac{\partial E}{\partial T^0_{i_0j_0k_0l_0}}=\sum_{i_nj_nk_nl_n}\frac{\partial E}{\partial  T^n_{i_nj_nk_nl_n}}
  \frac{\partial   T^n_{i_nj_nk_nl_n}}{\partial T^0_{i_0j_0k_0l_0}}.
\end{equation}
At each level we apply the chain rule repeatedly, leading to the form
\begin{eqnarray}
  \nonumber
  &&\frac{\partial E}{\partial T^{0}_{i_0j_0k_0l_0}}= \sum_{i_nj_nk_nl_n} \hskip-1.5mm \cdots \hskip-1.5mm \sum_{i_2j_2k_2l_2} \sum_{i_1j_1k_1l_1} 
  \frac{\partial E}{\partial T^n_{i_nj_nk_nl_n}} \times\\ 
  &&~~~~~~\frac{\partial T^{n}_{i_nj_nk_nl_n}}{\partial T^{n-1}_{i_{n-1}j_{n-1}k_{n-1}l_{n-1}}} 
{\cdots\frac{\partial T^2_{i_2j_2k_2l_2}}{\partial T^1_{i_1j_1k_1l_1}} \frac{\partial T^1_{i_1j_1k_1l_1}}{\partial T^0_{i_0j_0k_0l_0}}}.
\end{eqnarray}
All the factors can be easily computed and stored during the renormalization procedure, up to the highest level $n=n_{\rm max}$,  albeit at the rather high cost $\propto m^{24}$ 
operations when all contractions are carried out exactly (while with Monte Carlo sampling the cost is $m^{12}$).

We update the tensors, using the steepest decent method or a stochastic scheme were only the signs of the derivatives are used \cite{lou} (which to a large extent
avoids trapping in local minimas). In this process we normalize the elements of the $T^0(\sigma)$ and $S^n$ tensors such that the largest (in magnitude) elements in 
each are of order $1$ [in the case of $T^0(\sigma)$ taken as the largest among all elements for $\sigma=\pm 1$, since the $\sigma=\up$ and $\dn $ tensors cannot
be rescaled independently of each other). This can lead to problems with elements either too large or too small during the contraction of the full tensor network. 
At this stage, we therefore perform a separate rescaling in the following way: The individual $T^0(\sigma)$-tensors are normalized by contracting them on themselves 
as if they described a single isolated spin, i.e., first forming the double tensor $D_{abcd}$ in (\ref{decont}) and then contracting by summing $D_{aacc}$ over $a$ 
(assumed to correspond to the ``left'' and ``right'' bond indices) and $c$ (the ``up'' and ``down'' indices). The elements of $T^0(\sigma)$ are then rescaled so 
that this single-tensor contraction equals one. A similar rescaling is done for the $S^n$ tensors at each level, after contracting the renormalized plaquette tensors 
$T^n$ with themselves according to the above scheme. We have not encountered any numerical instabilities when this scheme is used.

It should be noted that the optimized $T$ tensors do not necessarily constitute a good TNS when contracted without the $S$ tensors, because they are optimized 
together. On the other hand, it should also be possible to construct special $S$ tensors that effectively perform something very similar to a SVD (although globally
optimized, not locally as in Ref.~\onlinecite{gu,xiang}) and then the optimized $T$ tensors by themselves should also form a good TNS when assembled into a standard 
2D tensor network. Here we do not impose any such conditions on the $S$ tensors.

\section{Results}
\label{sec3}

We now discuss calculations for the hamiltonian (\ref{ham}). For $L \to \infty$, the ground state of this system undergoes a quantum phase 
transition in the 3D Ising universality class, at a critical field $h_c/J\approx 3.04$ determined using quantum Monte Carlo simulations \cite{rieger}. Here we study 
the model using the smallest possible tensors and truncation, $m=m_{\rm cut}=2$, using the double-tensor approach for lattices of size $L=4,8,16,32$, and $64$. For the 
$T^0(\sigma)$ tensors, we enforce symmetry with respect to rotations of the indices, for a total of $6$ free parameters each for $T^0(1)$ and $T^0(-1)$. The $S$ 
tensors are symmetric in the two indices used in the internal contractions of the plaquettes, and so there are $6$ free parameters also for each $S^n$. 

Starting with random tensor elements, we calculate the energy and its derivatives with respect to all the parameters. We then update the parameters based on the 
derivatives, normally using a stochastic optimization scheme of the kind discussed in Ref.~\onlinecite{lou}. The plaquette renormalizations and the derivative calculations 
are highly parallelizable, which we take advantage of by using a massively parallel computer \cite{bluegene}. In general the method performs very well, although 
occasionally the energy converges to a local minimum, especially close to the critical point. 

\begin{figure}
\includegraphics[width=6.5cm, clip]{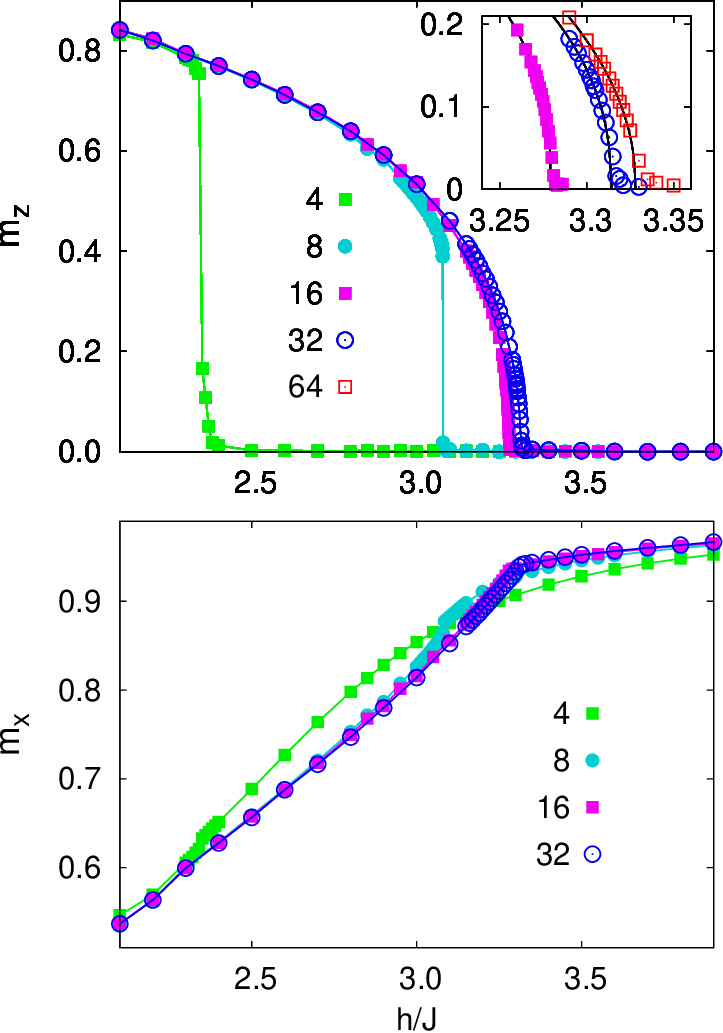}
\caption{(Color on-line) Spontaneous magnetization $m_z$ (upper panel) and field-induced polarization $m_x$ (lower panel) versus the transverse field for 
different $L\times L$ lattices. The inset in the upper panel shows the behavior close to the critical point in greater detail; the solid curves
are power-law fits, as discussed in the text and shown on a different scale in Fig.~\ref{crit}.}
\label{szsx}
\end{figure}

The method can produce solutions breaking spin-inversion symmetry, in which case we do not obtain the true ground state for a given system size $L$, which
can have no broken symmetries for finite $L$. This is to be expected, in analogy with how mean-field theory (corresponding to the $m=1$ simple product state) produces 
symmetry-broken states  for $h<h_c$. Thus we can simply compute the magnetization $m_z=\langle \sigma^z_s\rangle$ (averaged over all sites $s$) and study its behavior for 
increasing $L$. Note that for fixed $L$, spin-inversion symmetry should gradually be restored with increasing $m$, as the optimized state should approach the true finite-$L$ 
symmetric ground state. However, for any fixed $m$, we expect the symmetry to be broken when $L \to \infty$ for $h$ below some $m$-dependent $h_c$. Here we only 
consider $m=2$. Results for $m_z$ and the field-induced $m_x=\langle \sigma^x_s\rangle$ are shown versus the field $h/J$ in Fig.~\ref{szsx}. The transition point between 
the magnetic and paramagnetic states moves toward higher fields with increasing $L$, converging to $h_c/J \approx 3.33$ for the largest sizes. This is much
closer to the unbiased quantum Monte Carlo result \cite{rieger} $h_c/J\approx 3.04$ than the mean-field ($m=1$) value $h_c/J=4$. 

While other renromalization schemes
produce transition points even closer to the correct result for $m=2$ \cite{gu}, the results are not directly comparable because here we use $m_{\rm cut}=2$ throughout 
the renormalization, while much larger $m_{\rm cut}$ as used in \cite{gu}.

\begin{figure}
\includegraphics[width=7.5cm, clip]{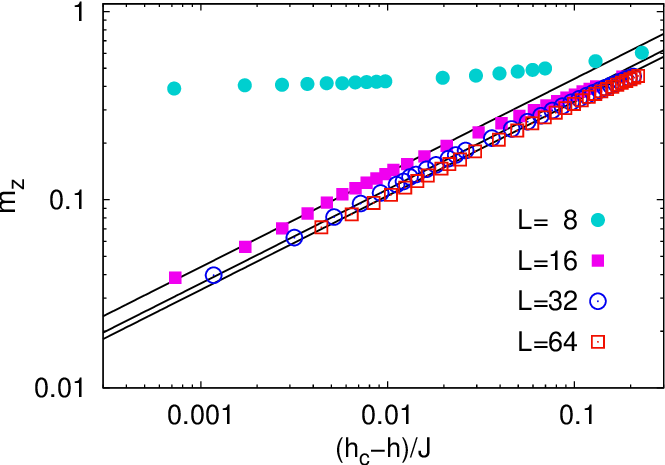}
\caption{(Color on-line) Critical scaling of the magnetization (the order parameter). The behavior for the larger system sizes is consistent
with mean-field behavior (exponent $\beta=1/2$), as shown with the straight lines.}
\label{crit}
\end{figure}

There is some rounding of the magnetization curve close to the transition for $h>h_c(L)$, which becomes less pronounced with increasing $L$. One the other hand, 
the magnetizaton curve for $h<h_c(L)$ becomes less sharp with increasing $L$. For $L=4$ and $8$ the transitions look almost first-order, with apparent jumps
of the order parameter which diminish with increasing $L$. This is very similar to the behavior found for matrix-product state with small matrices in
Ref.~\onlinecite{liu}, where it was shown that there is a crossing of two energy minimas in the space of matrix elements for finite systems. As the
system size grows, these minimas move close to each other and for infinite size the transition is continuous. If this mechanism is at play also here,
then there should be no finite size rounding associated with the jumps. The rounding that we do observe for $m^z$ in Fig.~\ref{szsx} can in principle be 
due to incomplete optimization of the tensors, which we cannot completely rule out. The results are, however, reproducible, and also the rounding effects
are smaller for the larger systems (where one would expect the optimization to be more difficult). It is therefore possible that the energy landscape
of the plaquette-renormalized TNS is different from that of the MPSs, and the jump is not completely discontinuous.

Analyzing the critical behavior by fitting a power-law, $(h_c-h)^\beta$, to the data for $L\ge 16$ (where we do not observe any jumps), we find that the behavior 
close to the transition can be described by a mean-field exponent, $\beta=1/2$, as shown in Fig.~\ref{crit}. For $L=8$ the behavior is different, with a very 
small exponent (related to the ``quasi-discontinuities'' discussed above). In Ref.~\onlinecite{liu} it was argued the critical behavior for TNSs should, in general, 
be expected to be of mean-field form for systems of infinite size and finite tensor (or matrix, in one dimension) size $m$. The true critical behavior (exponent)
should emerge with increasing $m$ (for large system sizes) in a window that tends toward $h_c$ as the system size is increased.

\begin{figure}
\includegraphics[width=7cm, clip]{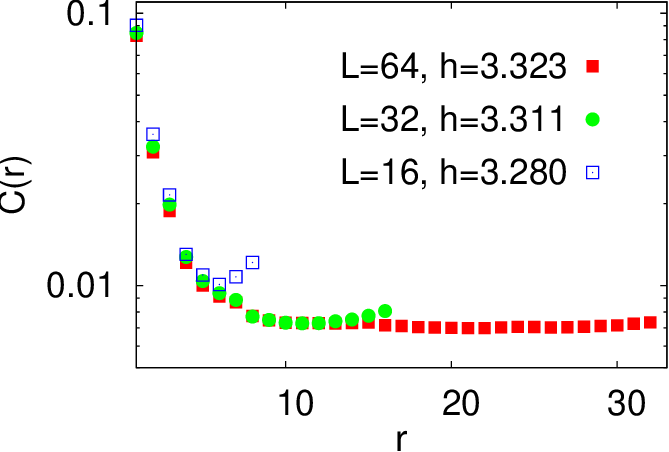}
\caption{(Color on-line) Spin-spin correlation versus separation $r$ at fields $h$ close to $h_c$ such that the long-distance 
correlations are approximately the same for system sizes $L=16,32$, and $64$.}
\label{corr}
\end{figure}

Fig.~\ref{corr} shows the spin-spin correlation function $C(r_{ij})=\langle \sigma^z_i\sigma^z_j\rangle$ averaged over all equidistant spins close to $h_c$. We have 
here chosen the particular values of $h$ for each system size in such a way that the curves for different system sizes coincide approximately and that the long-distance
value is relatively small (i.e., a weakly ordered system). The behavior is clearly different from a simple $m=1$ product state (mean-field theory), which only 
gives a constant $C(r)=\langle \sigma^z_i\rangle^2$ for $r>0$. The correlations instead decay over a distance of several lattice spacings. This shows that the
scheme can account for non-trivial quantum fluctuations. Characteristic boundary enhancements of the correlations at $r\approx L/2$ are also seen. It is not possible, 
however, to observe truly critical correlations with the small, $m=2$, tensors that we have used here. This is related to the asymptotic mean-field behavior seen 
in Fig.~\ref{crit}.

\begin{figure}
\includegraphics[width=8cm, clip]{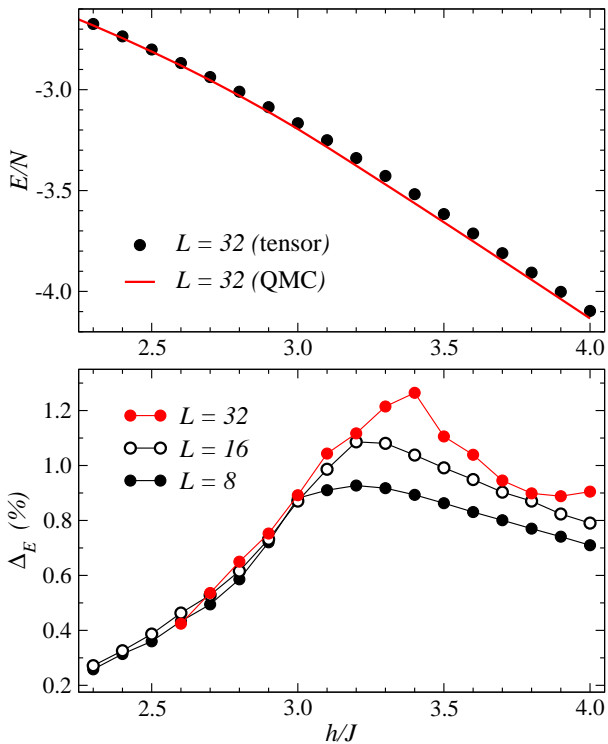}
\caption{(Color on-line) Upper panel: Ground state energy per spin obtained with the tensor-network ansatz for an $L=32$
lattice compared with unbiased quantum Monte Carlo calculations for the same systems. Lower panel: The relative energy error 
in the the tensor-network calculation (i.e., the deviation from the QMC results) for system sizes $L=8,16$, and $32$.}
\label{energy}
\end{figure}

We finally confirm that the calculations satisfy the variational bound of the energy. Unbiased results for comparison was obtained
using quantum Monte Carlo calculations \cite{aws03} for the same lattice sizes. Fig.~\ref{energy} shows the energy versus the
field for $L=32$, as well as the relative error for several system sizes. The results of the tensor-network calculations are
always higher. As expected, the error is peaked close to the phase transition. It also grows slightly with the system size, which
is also not unexpected. For $L=32$, some of the data points (the one at the peak as well as those for the largest fields) are 
likely affected by incomplete optimization of the tensors.

\section{Summary and discussion}
\label{sec4}

In summary, we have presented a scheme using auxiliary tensors to renormalize plaquette tensors in a 2D tensor network. The approach can also be regarded as a different 
tensor network, which can be contracted efficiently. Figs.~\ref{trg} and \ref{net2} summarize the approach pictorially. Using the example of the transverse-field Ising 
model, we have shown that the scheme produces results far better than mean-field theory, even with the smallest possible non-trivial tensors and truncation ($m=2$). 
The scaling to the thermodynamic limit is well-behaved. Based on these results, we expect a fast convergence to the exact ground state with increasing $m$. Increasing 
$m$ to $3$ is already quite challenging within the double-tensor approach that we have employed here, since the scaling is $\propto m^{24}$ (when computing all
energy derivatives). However, in variational Monte Carlo simulations (sampling the spins and optimizing the tensors based on stochastic estimates of the derivatives 
\cite{sandvik2}), the scaling is $m^{12}$, and it should then be easier to study larger $m$. Using some optimization method not requiring derivatives (e.g., the
methods discussed in \cite{liu}) formally brings the effort down to $m^8$ and  $m^{16}$ when using Monte Carlo sampling and exact spin tracing, 
respectively.

Here we optimized the tensors variationally, which requires the energy averaged over all non-equivalent sites and bonds. The computational effort then scales with the system 
size as $L^2\log(L)$. If we optimize only a local energy, which is not a variational estimate of the total energy but can produce good results in the SVD approach 
\cite{gu} (and may work well also in our scheme for some particular classes of $S$ tensors), the scaling is $\log(L)$, as in SVD based schemes. It may also be possible 
to use imaginary-time evolution (ground-state projection), as is often done in other TNS approaches \cite{schuch2}.

Application to other quantum spin systems (and even fermions) is in principle straight-forward, although the convergence with $m$ can of 
course be expected to be model dependent.

A technically appealing feature of our scheme is that the plaquette renormalization procedure can be implemented very efficiently on GPUs (graphics processing units, 
the use of which is emerging as an important trend in high-performance computing \cite{gpu}). The speed-up relative to a standard CPU can be very significant for large 
tensors. For $m=3$, we have achieved an efficiency boost of $\approx 25$ for the plaquette contraction of the double tensor in a more computationally challenging model, 
the $J_1-J_2$ Heisenberg model.\cite{GPUproceeding}. It should be noted that for the current case with $m=2$, the tensor size is too small to obtain any performance gain 
from GPU as the amount of computations needs to be done can not hide the memory latency. We plan to use this approach in future model studies with larger $m$.

We would like to thank Zheng-Cheng Gu, Frank Verstraete, and Xiao-Gang Wen for useful discussions. This work is supported by NSF Grant No.~DMR-0803510 (AWS) 
and by NSC Grant No.~97-2628-M-002-011-MY3 and NTU 99R0066-69 (YJK).  YJK. and AWS would also like to thank the NCTS of Taiwan for travel support.

\null

\end{document}